\documentclass[pra,aps,twocolumn,showpacs]{revtex4}

\usepackage{graphicx}
\usepackage[dvips]{color}
\usepackage{dcolumn}
\usepackage{amsmath}
\usepackage{amssymb}
\usepackage{latexsym}
\usepackage{bm}

\newcommand{\ket}[1]{\left| {#1} \right\rangle}
\newcommand{\bra}[1]{\left\langle {#1} \right|}

\begin{document}

\title{Fidelity of Quantum Teleportation through Noisy Channels}
\author{Sangchul Oh}
\email[Electronic address: ]{scoh@mrm.kaist.ac.kr}
\author{Soonchil Lee}
\email[Electronic address: ]{sclee@mail.kaist.ac.kr}
\author{Hai-woong Lee}
\email[Electronic address: ]{hwlee@laputa.kaist.ac.kr}
\affiliation{ Department of Physics,
              Korea Advanced Institute of Science and Technology,
              Daejon, 305-701, Korea}
\date{\today}

\begin{abstract}             
We investigate quantum teleportation through noisy quantum channels by 
solving analytically and numerically a master equation in the Lindblad 
form. We calculate the fidelity as a function of decoherence rates and 
angles of a state to be teleported.  It is found that the average 
fidelity and the range of states to be accurately teleported depend on 
types of noise acting on quantum channels. If the quantum channels is 
subject to isotropic noise, the average fidelity decays to $1/2$, which 
is smaller than the best possible value $2/3$ obtained only by the 
classical communication. On the other hand, if the noisy quantum 
channel is modeled by a single Lindblad operator, the average fidelity 
is always greater than $2/3$.
\end{abstract}
\maketitle

Quantum teleportation~\cite{Bennett93,Bouwmester97} is a process by 
which a sender, called Alice, transmits an unknown quantum state to 
a remote recipient, called Bob, via dual classical and quantum channels. 
Here a pair of maximally entangled particles, forming a quantum channel, 
should be used for the perfect quantum teleportation. However, while 
being distributed and kept by Alice and Bob, an entangled state may lose 
its coherence and become a mixed state due to the interaction with its 
environment.

Bennett {\it et al.}~\cite{Bennett93} noted that the quantum channel 
which is less entangled reduces the fidelity of teleportation, and/or 
the range of states that can be accurately teleported. 
Popescu~\cite{Popescu94} investigated the relations among teleportation, 
Bell's inequalities, and nonlocality.  It was demonstrated that there 
are mixed states which do not violate any Bell type inequality, but 
still can be used for teleportation. 
Horodecki {\it et al.}~\cite{Horodecki96} showed that any mixed two 
spin-$\frac{1}{2}$ state which violates the Bell-CHSH inequality is 
useful for teleportation.  Also Horodecki {\it et al.}~\cite{Horodecki99} 
proved the relation between the optimal fidelity of teleportation and 
the maximal singlet fraction of the quantum channel. 
Banaszek~\cite{Banaszek01} investigated the fidelity of quantum 
teleportation using non-maximally entangled states. 
Ishizaka~\cite{Ishizaka01} studied the quantum channel subject to local 
interaction with two-level environment. Although the studies cited above 
reveal the important relations between the degree of entanglement of the 
quantum channel and quantum teleportation, there seem to be little studies 
on the direct connection between the quantum teleportation and decoherence 
rates. Thus it might be interesting to know how the type and strength of 
noise acting on quantum channels affect the fidelity of quantum 
teleportation.

In this paper, we investigate quantum teleportation through noisy channels
by solving analytically and numerically a master equation in the Lindblad 
form. We obtain the fidelity of quantum teleportation as a function of 
decoherence time and angles of an unknown state to be teleported. Thus 
we explicitly demonstrate Bennett {\it et al.}'s argument that noisy 
quantum channels reduce the range of states to be accurately teleported. 
We also examine the characteristic dependence of the average fidelity on 
types of noise acting on qubits at each stage of the teleportation.

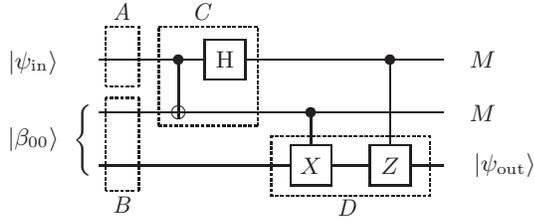
\begin{figure}[ht]
%
%
\begin{picture}(200,100)
\put(0,54){\makebox(10,10){$|\psi_{\text{in}}\rangle$}}
\put(0,25){\makebox(10,10){$|\beta_{00}\rangle$}}
\put(20,27.5){$\Biggl\{$}

\put(32.5,50){\dashbox{1}(12.5,22.5){}}
\put(52.5,35){\dashbox{1}(37.5,37.5){}}
\put(32.5,10.5){\dashbox{1}(12.5,35){}}
\put(95,8.5){\dashbox{1}(60,22.5){}}
\put(35,75){\it A}
\put(65,75){\it C}
\put(35,2){\it  B}
\put(120,1){\it D}

\put(30,60){\line(1,0){30}}
\put(60,60){\circle*{4}}
\put(60,60){\line(0,-1){22.5}}
\put(60,40){\circle{5}}
\put(60,60){\line(1,0){10}}
\put(70,52.5){\framebox(15,15){H}}
\put(85,60){\line(1,0){25}}

\put(30,40){\line(1,0){70}}
\put(30,20){\line(1,0){72.5}}
\put(110,40){\circle*{4}}
\put(110,40){\line(0,-1){12.5}}
\put(102.5,12.5){\framebox(15,15){$X$}}
\put(117.5,20){\line(1,0){14.5}}

\put(110,60){\line(1,0){30}}
\put(140,60){\circle*{4}}
\put(140,60){\line(0,-1){32.5}}
\put(132.5,12.5){\framebox(15,15){$Z$}}

\put(100,40){\line(1,0){60}}
\put(147.5,20){\line(1,0){12.5}}
\put(100,60){\line(1,0){60}}
\put(170,55){\makebox(10,10){$M$}}
\put(170,35){\makebox(10,10){$M$}}
\put(180,15){\makebox(10,10){$|\psi_{\text{out}}\rangle$ }}
\end{picture}
\caption{\label{fig:teleport.cicuit}
   A circuit for quantum teleportation through noisy channels. 
   The two top lines belong to Alice, while the bottom one to Bob. 
   $M$ represents measurement. The dotted boxes, {\it A}, 
   {\it B}, {\it C}, and {\it D} denote noisy channels. 
   Time advances from left to right. During the time 
   interval corresponding to the width of the dotted box, 
   the Lindblad operator is turned on.}
\end{figure}

Let us consider quantum teleportation through noisy channels as 
illustrated in Fig.~\ref{fig:teleport.cicuit}. The top two qubits are 
taken by Alice and the bottom qubit is kept by Bob. Here measurements 
are performed at the end of the circuit for computational convenience. 
Classical conditional operations can be replaced with corresponding 
quantum conditional operations~\cite{Nielsen00}. Decoherence of an open 
quantum system is due to the interaction with its environment. Under 
the assumption of Markov and Born approximations and after tracing out 
the environment degrees of freedom, the dynamics of an open quantum system 
is described by a master equation for the density operator of the quantum 
system alone, $\rho(t)$ in the Lindblad form~\cite{Lind76,Alicki87} \begin{eqnarray}
\frac{\partial\rho}{\partial t} 
= -\frac{i}{\hbar} [H_S,\rho]
+ \sum_{i,\alpha}\Bigl(  L_{i,\alpha}\rho L_{i,\alpha}^\dagger 
   - \frac{1}{2}\{L_{i,\alpha}^\dagger L_{i,\alpha}, \rho\}\Bigr),
\label{Eq:Lindblad}
\end{eqnarray}
where the Lindblad operator 
$L_{i,\alpha} = \sqrt{\kappa_{i,\alpha}(t)}\,\sigma_{\alpha}^{(i)}$ 
acts on the $i-$th qubit and describes decoherence.
Throughout this paper, $\sigma^{(i)}_\alpha$ denotes the Pauli spin matrix
of the $i$-th qubit with $\alpha = x,y,z$. The decoherence time is 
approximately given by $1/\kappa_{i,\alpha}$. By switching on and off 
$\kappa_{i,\alpha}(t)$ we could control noise. 
We take the Hamiltonian of a qubit system as an ideal model of a quantum 
computer which is given by~\cite{Makhlin01}
\begin{eqnarray}
 H_S(t) = -\frac{1}{2}\sum_{i=1}^N
             \mathbf{B}^{(i)}(t)\bm{\cdot}\bm{\sigma}^{(i)}
        -\sum_{i\ne j}J_{ij}(t)\sigma_+^{(i)}\sigma_-^{(j)},
\label{Eq:Hamiltonian}
\end{eqnarray}
where $\bm{\sigma}^{(i)} = (\sigma_x^{(i)},\sigma_y^{(i)},\sigma_z^{(i)})$ 
and $\sigma_\pm^{(i)} = \frac{1}{2}(\sigma_x^{(i)} \pm i\sigma_y^{(i)})$.
In solid state qubits, various types of the coupling between qubit $i$ 
and $j$ are possible such as the $XY$ coupling given above, the Heisenberg 
coupling, and the Ising coupling $J_{ij}\sigma_z^{(i)}\sigma_z^{(j)}$ in NMR. 
The various quantum gates in Fig.~\ref{fig:teleport.cicuit} could be 
implemented by a sequence of pulses, i.e., by turning on and off 
$\mathbf{B}^{(i)}(t)$ and $J_{ij}(t)$. We develop the simulation code which 
solves Eq.~(\ref{Eq:Lindblad}), the set of differential equations for 
the density matrix $\rho_{mn}(t)$, based on the Runge-Kutta 
method~\cite{Oh02}. Eq.~(\ref{Eq:Lindblad}) shows 
$\text{Tr}\rho(t) = 1$ at all times. 

An unknown state to be teleported can be written as
$|\psi_{\rm in}\rangle = \alpha |0\rangle + \beta |1\rangle$ with 
$|\alpha|^2 + |\beta|^2 =1$. It is convenient to rewrite 
$|\psi_{\rm in}\rangle$ as a Bloch vector  on a Bloch sphere
\begin{eqnarray}
|\psi_{\rm in}\rangle 
     = \cos\biggl(\frac{\theta}{2}\biggr) e^{ i\phi/2} |0\rangle 
     + \sin\biggl(\frac{\theta}{2}\biggr) e^{-i\phi/2} |1\rangle,
\end{eqnarray}
where $\theta$ and $\phi$ are the polar and azimuthal angles, respectively.
The maximally entangled state of two spin-$\frac{1}{2}$ particles shared 
and kept by Alice and Bob is given by
\begin{eqnarray}
|\beta_{00}\rangle \equiv \frac{1}{\sqrt{2}}( |00\rangle + |11\rangle).
\end{eqnarray}
The input state of the quantum teleportation circuit in 
Fig.~\ref{fig:teleport.cicuit} is the product state of $|\psi_{\rm in}\rangle$ 
and $|\beta_{00}\rangle$. After the implementation of the quantum circuit of
Fig.~\ref{fig:teleport.cicuit} and the measurement of the top two qubits,
Bob gets the teleported state $|\psi_{\rm out}\rangle$. It is useful to 
describe the teleportation in terms of density operators
\begin{eqnarray}
\rho_{\rm out} 
&=&\text{Tr}_{1,2}\bigl\{
    U_{\text{tel}}\, \rho_{\text{in}}\otimes\rho_{\text{en}}\,
    U_{\text{tel}}^\dagger \bigr\},
\label{Eq:teleport}
\end{eqnarray}
where $\rho_{\text{in}} =\ket{\psi_{\text{in}}}\bra{\psi_{\text{in}}}$,
$\rho_{\text{en}} =\ket{\beta_{00}}\bra{\beta_{00}}$, and 
$\text{Tr}_{1,2}$ is a partial trace over qubits 1 and 2.
The unitary operator $U_{\text{tel}}$ is implemented by the teleportation 
circuit as shown in Fig.~\ref{fig:teleport.cicuit}. If the teleportation 
is ideal, the density matrix teleported $\rho_{\text{out}}$ is identical to 
$\rho_{\text{in}}$ up to the normalization factor.

As illustrated as dotted boxes in Fig.~\ref{fig:teleport.cicuit}, 
we consider four different noisy channels, {\it A, B, C}, and {\it D}. 
In case {\it A} an unknown state $|\psi_{\text{in}}\rangle$ loses its 
coherence and becomes a mixed state before it is teleported. In case 
{\it B} an entangled pair, forming a quantum channel, become noisy 
while being shared and kept by Alice and Bob.  In cases {\it C} and 
{\it D}, while Alice and Bob perform the Bell measurement and the unitary 
operation, respectively, noise may set in. For cases {\it A} and 
{\it B} we obtain both analytic and numerical solutions of 
Eq.~(\ref{Eq:Lindblad}). While in cases {\it C} and {\it D} the numerical 
solutions of Eq.~(\ref{Eq:Lindblad}) are obtained.  For our numerical 
calculation, $\kappa_{i,\alpha}(t)$ is turned on for the time interval 
$\tau$ corresponding to the width of each dotted boxes in 
Fig.~\ref{fig:teleport.cicuit}.
 
The properties of quantum teleportation through noisy quantum channels 
are quantified by the fidelity which measures the overlap between 
a state to be teleported $|\psi_{\text{in}}\rangle$ and the density 
operator for a teleported state $\rho_{\text{out}}$
\begin{eqnarray}
F(\theta,\phi) 
= \langle\psi_{\rm in}|\rho_{\rm out}|\psi_{\text{in}}\rangle.
\label{Eq:fidelity}
\end{eqnarray}
Here the fidelity $F(\theta,\phi)$ depends on an input state as well 
as the type of noise acting on qubits. We calculate $F(\theta,\phi)$ 
and determine the range of states $|\psi_{\text{in}}\rangle$ which can 
be accurately teleported. Since in general a state to be teleported is 
unknown, it is more useful to calculate the average fidelity given by
\begin{eqnarray}
F_{\text{av}} = \frac{1}{4\pi}\int_0^{\pi}d\theta\int_0^{2\pi}d\phi\, 
F(\theta,\phi)\sin\theta\,,
\label{def_avg_fidelity}
\end{eqnarray}
where $4\pi$ is the solid angle.
\paragraph*{\it Case A: states to be teleported are mixed.} 
Alice is not able to know or copy the state to be teleported without 
disturbing it. So it may be pure or mixed.  As Bennett 
{\it et al.}~\cite{Bennett93} noted, the linear property of quantum 
teleportation enables one to teleport not only a pure state but also 
a mixed state. The quantum operation ${\cal E}$ transforms a pure state 
$\rho_{\text{in}}=\ket{\psi_{\text{in}}}\bra{\psi_{\text{in}}}$ to a mixed state 
${\cal E}(\rho_{\text{in}})$. The time-evolution of pure states to mixed 
states is described by Eq.~(\ref{Eq:Lindblad}). 
See Ref.~\onlinecite{Nielsen00} for the connection between two approaches. 
From Eq.~(\ref{Eq:teleport}), quantum teleportation of mixed states reads
\begin{eqnarray}
{\cal E}(\rho_{\text{out}}) &=& \text{Tr}_{1,2}\bigl\{
             U_{\text{tel}}\, {\cal E}(\rho_{\text{in}})\otimes\rho_{\text{en}}\,
             U_{\text{tel}}^\dagger \bigr\}\,.
\label{Eq:Case-A}
\end{eqnarray}
The decoherence of the state to be teleported, ${\cal E}(\rho_{\text{in}})$ 
is transfered to the state teleported, ${\cal E}(\rho_{\text{out}})$.
For various types of noise, we obtain both analytic and numerical 
solutions of Eq.~(\ref{Eq:Lindblad}), and calculate the fidelity.

\begin{figure}
\includegraphics[]{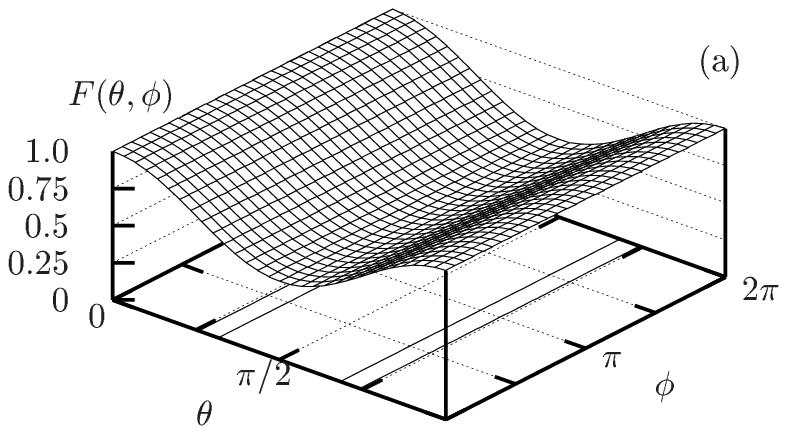}\\[12pt]
\includegraphics[]{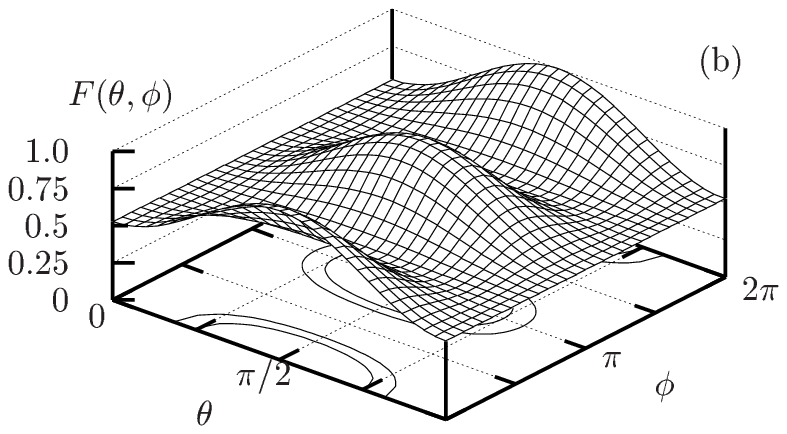}\\[12pt]
\includegraphics[]{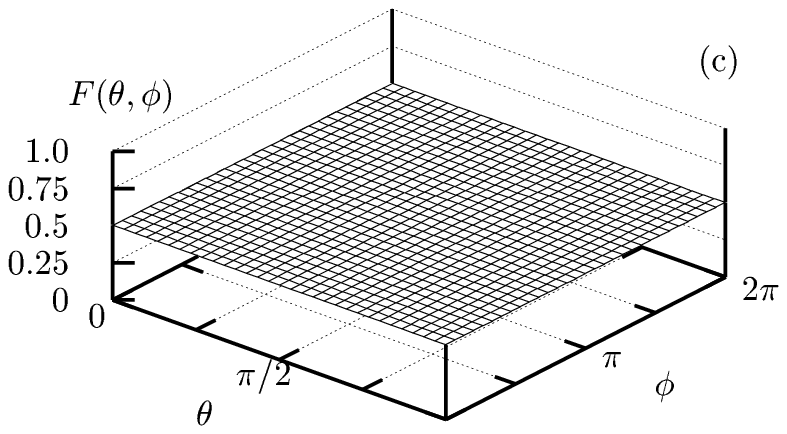}
\caption{\label{fig:Case-AB-fid}
   Fidelity $F(\theta,\phi)$ as a function of 
   angles $\theta$ and $\phi$ of the state to be teleported 
   for case {\it A} at $2\kappa \tau = 3.0$ and for case {\it B} at 
   $4\kappa \tau = 3.0$. For case {\it B} in (a) the Lindblad operators 
   $L_{2,z}$ and $L_{3,z}$, and in (b) $L_{2,x}$ and $L_{3,x}$ are turned 
   on. The maximum values of the fidelity of (a) and (b) are 1 and 
   the minimum $1/2$. The contours on the $\theta-\phi$ planes
   in (a) and (b) join the points with the fidelity $F(\theta,\phi)=3/4$ 
   and $2/3$, respectively. In (c) the isotropic noise is applied.}
\end{figure}

Suppose a state to be teleport is subject to the noise $L_{1,z}$. It is 
easy to find the analytic solution of Eq.~(\ref{Eq:Lindblad}) when 
$H_S(t)=0$. We obtain the mixed state to be teleported, 
${\cal E}(\rho_{\text{in}})$ as $\rho^{(00)}(t)=\rho^{(00)}_{\text{in}}(0)$, 
$\rho^{(11)}(t) = \rho^{(11)}_{\text{in}}(0)$, 
and $\rho^{(01)}(t) = \rho^{(01)}_{\text{in}}(0) \exp(-2\kappa t)$. 
Then from Eqs.~(\ref{Eq:Case-A}) and~(\ref{Eq:fidelity}), the fidelity 
can be calculated as
\begin{eqnarray}
F(\theta,\phi) = 1 -\frac{1}{2}(1 - e^{-2\kappa\tau}) \sin^2\theta\,.
\label{Eq:fidel_Lz1}
\end{eqnarray}
If $2\kappa\tau \ll 1$, $F(\theta,\phi)\simeq 1 -\kappa\tau\sin^2\theta$.
On the other hand, if $2\kappa\tau \gg 1$, 
$F(\theta,\phi)\simeq \frac{1}{2}(1 + \cos^2\theta)$.
Fig.~\ref{fig:Case-AB-fid}~(a) is the plot of Eq. (\ref{Eq:fidel_Lz1})
for $2\kappa\tau = 3.0$.

Let us consider the state $\ket{\psi_{\rm in}}$ is subject to the noise
described by $L_{1,x}$. After some calculations, we obtain the fidelity
\begin{eqnarray}
 F(\theta,\phi) 
 &=& \frac{1}{2}\bigl[ 1 + \sin^2\theta\cos^2\phi \nonumber \\
 &+& e^{-2\kappa\tau}( \cos^2\theta +\sin^2\theta\sin^2\phi) \bigr] \,.
 \label{Eq:fidel_Lx1}
\end{eqnarray}
If $2\kappa\tau \ll 1$, $F(\theta,\phi) \simeq 1 
- \kappa\tau\Bigl( \cos^2\theta + \sin^2\theta\sin^2\phi\Bigr)$.
In the limit of $2\kappa\tau \gg 1$, we have 
$F(\theta,\phi) \simeq \frac{1}{2}(1 + \sin^2\theta\cos^2\phi)$.
The plot of Eq. (\ref{Eq:fidel_Lx1}) at $2\kappa\tau = 3.0$ is shown in
Fig.~\ref{fig:Case-AB-fid}~(b).  

\begin{figure}[htbp]
\includegraphics[width =8cm]{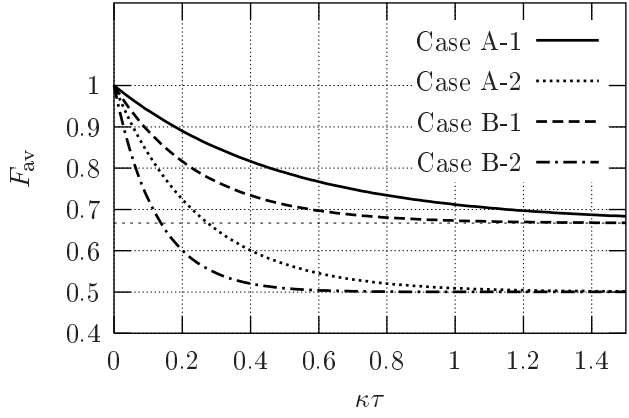}
\caption{\label{fig:Case-AB-avgfid}
   Average fidelity $F_{\rm av}$ as a function of 
   $\kappa\tau$ for cases {\it A} and {\it B}.
   The solid line (Case A-1) is the plot of 
   Eq.~(\ref{Eq:Case-A1}) for the noise described by $L_{1,x}$ 
   (or by $L_{1,z}$). The dotted line (Case A-2) 
   is based on Eq.~(\ref{Eq:Case-A2}) corresponding to the isotropic noise. 
   The dashed line (Case B-1) is for Eq.~(\ref{Eq:Case-B1}), 
   the noise modeled by $L_{2,x}$ and $L_{3,x}$ (or by $L_{2,z}$ and $L_{3,z}$). 
   The dash-dotted line (Case B-2) is the plot of Eq.~(\ref{Eq:Case-B2})
   for the isotropic noise. The horizontal dotted line with $2/3$ shows 
   the maximum fidelity obtained only by the classical communication.}
\end{figure}

Substituting Eqs.~(\ref{Eq:fidel_Lx1}) or~(\ref{Eq:fidel_Lz1}) into 
Eq.~(\ref{def_avg_fidelity}), we get the average fidelity 
\begin{eqnarray}
F_{\text{av}}(\tau) = \frac{2}{3} + \frac{1}{3}\,e^{-2\kappa \tau}.
\label{Eq:Case-A1}
\end{eqnarray}
In Fig.~\ref{fig:Case-AB-avgfid}, the solid line (denoted by Case A-1) 
shows the plot of Eq.~(\ref{Eq:Case-A1}), the average fidelity as a function 
of $\kappa\tau$ for the noise modeled by $L_{1x}$ or $L_{1z}$.

Now suppose the isotropic noise ($L_{1x}, L_{1y}$, and $L_{1z}$) is applied 
to the state $\ket{\psi_{\rm in}}$. The analytic solution of 
Eq.~(\ref{Eq:Lindblad}) gives us the fidelity written by
\begin{eqnarray}
F_{\text{av}} = F(\theta,\phi) 
            = \frac{1}{2} + \frac{1}{2}\,e^{-4\kappa \tau}\,.
\label{Eq:Case-A2}
\end{eqnarray}
If $4\kappa\tau \ll 1$, $F(\theta,\phi)\simeq 1 -2\kappa\tau$.
For $4\kappa\tau \gg 1$, we have $F(\theta,\phi)\simeq \frac{1}{2}$
as shown in Fig.~\ref{fig:Case-AB-fid}~(c). 
In Fig.~\ref{fig:Case-AB-avgfid}, the dotted line (denoted by Case A-2) is
the plot of Eq.~(\ref{Eq:Case-A2}).
\paragraph*{Case B: Quantum channels are noisy.} 
While being distributed and stored by Alice and Bob, 
an entangled state of two spin-$\frac{1}{2}$ particles may 
be subject to noise. The dynamics of an entangled pair subject to
quantum noise is described by the quantum operation ${\cal E}$ 
acting on the pure entangled state, 
$\rho_{\text{en}} \to {\cal E}(\rho_{\text{en}})$
or by Eq.~(\ref{Eq:Lindblad}). From Eq.~(\ref{Eq:teleport}), 
the quantum teleportation with noisy quantum channels can be written as
\begin{eqnarray}
{\cal E}(\rho_{\text{out}}) 
  = \text{Tr}_{1,2}\bigl\{
     U_{\text{tel}}\, \rho_{\text{in}}\otimes{\cal E}(\rho_{\text{en}})\,
     U_{\text{tel}}^\dagger \bigr\}\,.
\end{eqnarray}
We find that the quantum teleportation process transfers the decoherence of 
the entangled pair ${\cal E}(\rho_{\text{en}})$ to that of the output state
${\cal E}(\rho_{\text{out}})$. It should be noted that the quantum operation 
acting on the entangled pair ${\cal E}(\rho_{\text{en}})$ is a $4\times4$ 
matrix but effectively a $2\times2$ matrix.  Thus overall features of 
case {\it B} are similar to case {\it A} except decoherence rates.

Consider the quantum channel subject to the noise acting in one direction, 
for example, the $z$ direction. This type of noise is modeled by Lindblad 
operators, $L_{2,z}=\sqrt{\kappa_{2,z}}\,\sigma^{(2)}_z$ and 
$L_{3,z}=\sqrt{\kappa_{3,z}}\,\sigma^{(3)}_z$, acting on an entangled pair, 
qubit 2 and qubit 3, respectively. Here we assume the same strength of 
decoherence rates, $\kappa\equiv\kappa_{2,z} = \kappa_{3,z}$. 
We obtain the fidelity $F(\theta,\phi)$ with the same form of 
Eq.~(\ref{Eq:fidel_Lz1}) except the replacement of 
$2\kappa\tau$ with $4\kappa\tau$. That is  
$F(\theta,\phi) = 1 -\frac{1}{2}[1 - \exp(-4\kappa\tau)]\sin^2\theta$.
For the noise described by $L_{2,x}$ and
$L_{3,x}$, the fidelity $F(\theta,\phi)$ is identical to 
the form of Eq.~(\ref{Eq:fidel_Lx1}) with exponent $4\kappa\tau$.

Let us discuss Bennett {\it et al.}'s argument: the imperfect quantum channel 
reduces the range of state $\ket{\psi_{\rm in}}$ that is accurately 
teleported~\cite{Bennett93}. Fig.~\ref{fig:Case-AB-fid} shows 
the fidelity $F(\theta,\phi)$ for various types of noise at 
$4\kappa \tau = 3.0$. For the noisy channel defined by $L_{2,z}$ 
and $L_{3,z}$, the fidelity $F(\theta,\phi)$ is always the maximum value 
1 at $\theta = 0, \pi$ irrespective of $\kappa\tau$ as depicted 
in Fig.~\ref{fig:Case-AB-fid}~(a).  These angles indicate states 
$|0\rangle$ and $|1\rangle$, which are eigenstates of $\sigma_z$. 
From $F(\theta,\phi)\simeq\frac{1}{2}(1+\cos^2\theta)$ in the limit of 
$4\kappa\tau \gg 1$, the range of states to be teleported with fidelity 
$F\ge 3/4$ is determined by $0\le \theta \le {\pi/4}$ and 
${3\pi/4}\le \theta \le\pi$. The teleported states with fidelity 
$2/3$ are in the region determined by $\cos\theta\ge {1/\sqrt{3}}$ 
or $\cos\theta\le -{1/\sqrt{3}}$. When $L_{2,x}$ and $L_{3,x}$ are 
applied to the qubits 2 and 3, we get $F(\theta,\phi)=1$ at $\theta = \pi/2$ 
and $\phi = 0, \pi$ for $4\kappa\tau  \gg 1$, which shown in 
Fig.~\ref{fig:Case-AB-fid}~(b). These angles represent states 
$\ket{\psi}=\frac{1}{\sqrt{2}}(\ket{0} + \ket{1})$ and 
$\ket{\psi}=\frac{1}{\sqrt{2}}(\ket{0} - \ket{1})$, i.e.,
eigenstates of $\sigma_x$.  The range of states accurately teleported 
is depicted by contours in Fig.~\ref{fig:Case-AB-fid}~(a) and (b).

When the quantum channel is subject to noise in an one direction, we obtain 
the average fidelity as depicted in Fig.~\ref{fig:Case-AB-avgfid} 
(denoted by Case B-1)
\begin{eqnarray}
F_{\text{av}}(\tau) = \frac{2}{3} + \frac{1}{3}\,e^{-4\kappa \tau}\,.
\label{Eq:Case-B1}
\end{eqnarray}
The average fidelity decays exponentially to the limiting value of $2/3$. 
This is the best possible score when Alice and Bob 
communicate each other only through the classical 
channel~\cite{Popescu94,Massar95}. 

Consider the case that the quantum channel is affected 
by isotropic noise, which is described by six Lindblad operators,
$L_{2,\alpha}$ and $L_{3,\alpha}$ with $\alpha = x,y,z$.
Then the analytic calculation of the fidelity can be written by 
\begin{eqnarray}
F_{\text{av}} = F(\theta,\phi)= \frac{1}{2} + \frac{1}{2}\,e^{-8\kappa \tau}.
\label{Eq:Case-B2}
\end{eqnarray}
As depicted in Fig.~\ref{fig:Case-AB-fid}~(c), the fidelity $F(\theta,\phi)$ 
is independent of angles of input states, $\theta$ and $\phi$ 
for any value of $\kappa\tau$. For the quantum channel subject to isotropic noise,
one could not find the range of states that is accurately teleported.
As shown in Fig.~\ref{fig:Case-AB-avgfid} (Case B-2), 
the average fidelity decay to the value $1/2$.
The number $1/2$ can be obtained when Alice and Bob
can not communicate at all and Bob merely selects a state 
at random. 

It should be noted that except decoherence rates $\kappa\tau$, 
the overall features of cases {\it A} and {\it B} are identical. 
This implies that if a state to be teleported is realized by
a single particle not an ensemble, one may not be able to identify 
whether the state to be teleported is mixed or 
the quantum channel is noisy.  

\paragraph*{Cases C and D: Noise during Bell's measurement 
or the unitary operation.} 
When Alice performs the Bell's measurement or Bob does the unitary operation 
on his particle of an entangled pair, noise may take place as 
depicted by the boxes {\it C} or {\it D} in Fig.~\ref{fig:teleport.cicuit}. 
In contrast to cases {\it A} and {\it B}, it seems to be difficult to find 
analytic solutions of Eq.~(\ref{Eq:Lindblad}) for cases {\it C} and {\it D} 
because of the time-dependence of the qubit Hamiltonian $H_S(t)$. 
Alice's Bell measurement on qubits 1 and 2 could be done by a controlled-not 
gate (CNOT) on qubits 1 and 2, and a Hadamard gate H$_1$ on qubit 1 as shown 
in Fig.~\ref{fig:teleport.cicuit}. With a qubit system modeled by Hamiltonian
Eq.~(\ref{Eq:Hamiltonian}), the CNOT gate acting on qubits 1 and 2 could 
be implemented by the pulse sequence~\cite{Makhlin01,Oh02}
$\text{CNOT} = e^{-i\pi/4}\text{H}_1 R_{2x}(\frac{\pi}{2}) R_{1x}(-\frac{\pi}{2}) 
 U_{2b}^{12}(\frac{\pi}{4}) R_{1x}(\pi) U_{2b}^{12}(\frac{\pi}{4}) \text{H}_1$.
Here $R_{jx}(\theta) \equiv e^{i\sigma_x^{(j)}\theta/2}$ is a rotation
of qubit $j$ by angle $\theta$ about the $x$ axis. A two qubit operation 
$U_{2b}^{12}(\theta)$ on qubits 1 and 2 is implemented by turning on the 
coupling $J_{12}$ for a time $t$ corresponding to $\theta\equiv J_{12}t/\hbar$.
During each qubit operation, the noise modeled by Lindblad operators is 
also switched on. Thus it seems to be not simple to obtain an analytic solution
and we takes a numerical method to solve the problem.

\begin{figure}[htbp]
\includegraphics[width = 8cm]{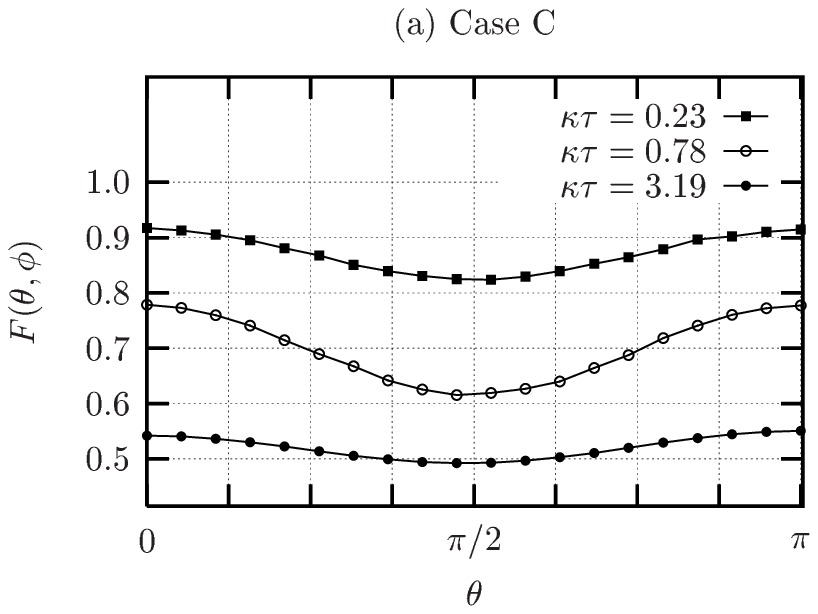}\\[12pt]
\includegraphics[width = 8cm]{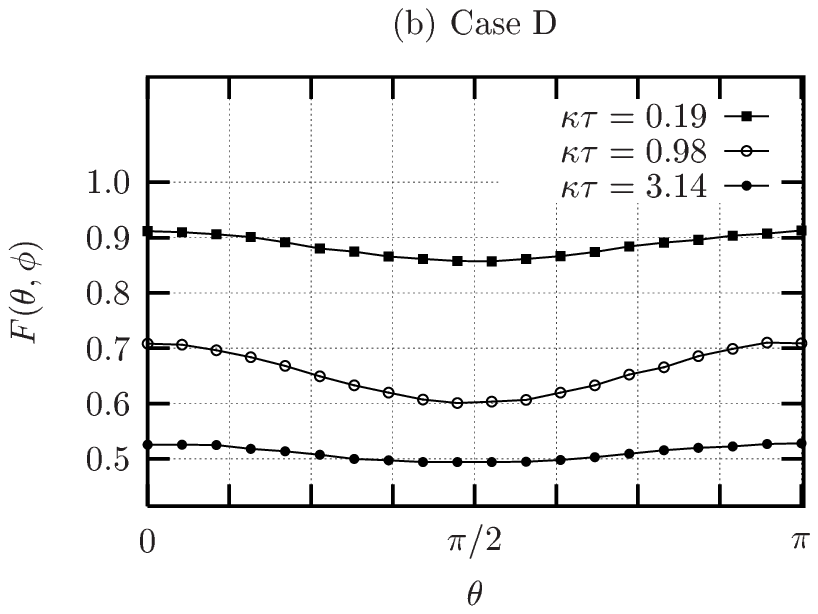}\\[12pt]
\includegraphics[width = 8cm]{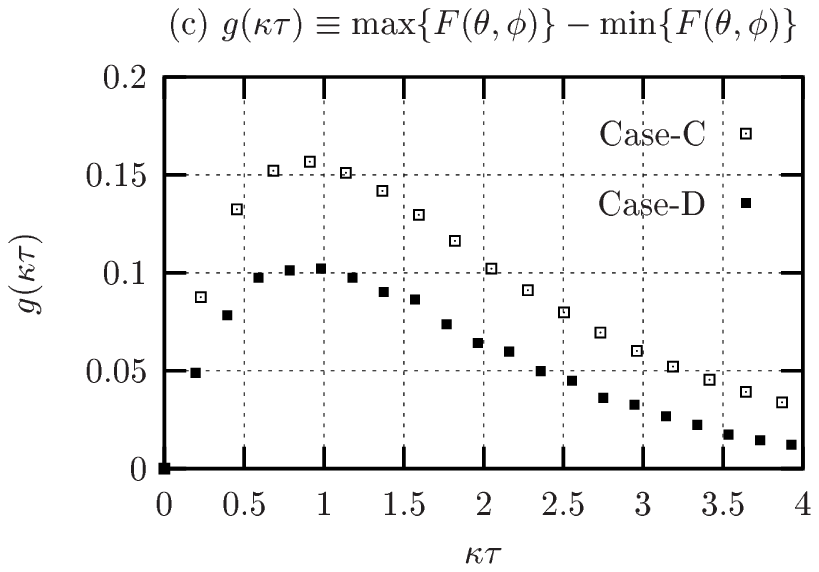}
\caption{\label{fig:Case-CD-fid}
   Fidelity $F(\theta,\phi)$ vs angle $\theta$ for various values 
   of $\kappa\tau$. (a) for case {\it C} noise is modeled by $L_{1z}$ 
   and $L_{2z}$, and (b) for case {\it D} by $L_{3z}$. $F(\theta,\phi)$ 
   is independent of angle $\phi$ because of the cylindrical symmetry 
   of $L_{iz}$ with $i=1,2,3$. Differences between the maximum and 
   minimum values of the fidelity $F(\theta,\phi)$ are plotted 
   as a function of $\kappa\tau$.}
\end{figure}

\begin{figure}[htbp]
\includegraphics[width = 8cm, scale = 1.0]{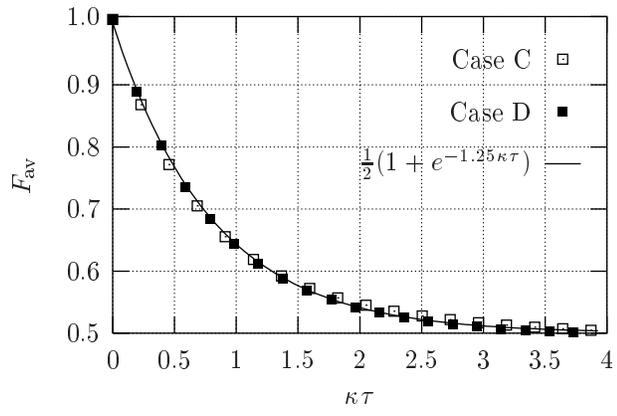}
\caption{\label{fig:Case-CD-avgfid}
   Average fidelity $F_{\rm av}$ as a function of $\kappa \tau$ 
   for cases {\it C} (dotted boxes) and {\it D} (filled boxes).}
\end{figure}

Consider the noise modeled by the Lindblad operators, 
$L_{1z}$ and $L_{2z}$ for case {\it C}, and $L_{3z}$ for case {\it D}.
Here the noise is switched on during 
the time interval $\tau$ corresponding to 
the total operation time which it takes to implement Bell's 
measurement or controlled $X$ and $Z$ operations. 
The time interval $\tau$ depends on the operation times of 
a single gate or a two qubit gate, proportional to $h/|{\bf B}^{(i)}|$ and 
$h/J_{ij}$, respectively.
Fig.~\ref{fig:Case-CD-fid} shows the fidelity $F(\theta,\phi)$
as a function of angle $\theta$ for various values of $\kappa\tau$. 
In contrast to the previous cases (Case A-1, Case A-2, Case B-1, 
and Case B-2) whose fidelity is given by Eqs.~(\ref{Eq:fidel_Lz1}) 
or~(\ref{Eq:fidel_Lx1}), in cases {\it C} and {\it D} the degrees of 
the dependence of fidelity $F(\theta,\phi)$ on angles $\theta$ 
is maximum at a certain value of $\kappa\tau$. 
Fig.~\ref{fig:Case-CD-fid} (c) shows the differences between the maximum 
and minimum values of the fidelity, 
$g(\kappa\tau)\equiv \text{max}\{F(\theta,\phi)\} - \text{min}\{F(\theta,\phi)\}$.
It is not clear why $g(\kappa\tau)$ has the maximum at $\kappa\tau\approx 0.98$.
As depicted in Fig.~\ref{fig:Case-CD-avgfid}, 
the average fidelity falls to the value $1/2$ and is approximately fitted by
\begin{eqnarray}
F_{\text{av}}(\tau) = \frac{1}{2} + \frac{1}{2}\,e^{-1.25\kappa \tau}\,.
\end{eqnarray}
One sees that cases {\it A} and {\it B} entirely differ from 
cases {\it C} and {\it D}.
Although an analytic solution for case {\it C} and {\it D} can not be 
obtained, it can be understood why the average fidelity decays to $1/2$ 
despite noise described by the Lindblad operator acting in one direction. 
Consider a rotation of a qubit about the $x$ axis in the presence of 
noise modeled by $L_z$. A simple calculation shows that the Bloch 
vector ${\bf r}$ of a qubit, which is defined by
$\rho = \frac{1}{2} (1 + {\bf r}{\bm\cdot \bm\sigma})$, falls to zero for 
any initial state. This means the qubit is depolarized and becomes 
a totally mixed state. Thus the average fidelity decays to $1/2$ when a
gate operation is done in the presence of noise.

It is valuable to discuss our results in connection with the previous
studies~\cite{Popescu94}. In Ref.~\onlinecite{Popescu94}, 
Popescu illustrated an example of a mixed pair which does not violate 
any Bell inequality but has an average fidelity $3/4$ for arbitrary input 
states, given by $\rho =\frac{1}{8}I + \frac{1}{2}\ket{\Psi^-}\bra{\Psi^-}$ 
with $\ket{\Psi^-} = \frac{1}{\sqrt{2}}(\ket{01} - \ket{10})$. 
Our calculation in which the quantum channel is described by 
this mixed state shows the fidelity $F(\theta,\phi)={3}/{4}$, 
independent of angles $\theta$ and $\phi$ and thus gives us an average 
fidelity $F_{\text{av}}=3/4$. 
Horodecki {\it et. al.}~\cite{Horodecki99} showed that the optimal 
fidelity of the standard quantum teleportation is given by
$f = (2F_{AB} + 1)/3$, where $F_{AB}$ is the singlet fraction of 
the quantum channel. From Eq.~(\ref{Eq:Case-B1}), one can write 
$F_{AB}= (1 + e^{-4\kappa\tau})/2$.

In conclusion, we calculated the fidelity and the average fidelity of 
quantum teleportation subject to various types of noise during different 
steps of the teleportation.  We examined the range of states that can 
be accurately teleported. Among states to be teleported, the eigenstate 
of the Lindblad operators is less sensitive to the noise.  It was 
shown that one can not distinguish whether an unknown state to be 
teleported, which is realized by a single particle, is mixed or the 
quantum channel is noisy. We found the dependence of the average 
fidelity on the type of noise affecting the quantum channel. 
If the quantum channel is subject to isotropic noise, the average 
fidelity may decay to $1/2$. On the other hand, if the noisy quantum 
channel is described by a single Lindblad operator, the average fidelity 
is always greater than the value $2/3$, the best possible value which 
can be obtained only by the classical communication.

\begin{acknowledgments}
We thank Dr. Jaewan Kim for helpful discussions. This work was supported 
by the Brain Korea 21 project of the Korea Ministry of Education, the NRL 
program of the Korean Ministry of Science and Technology, and KOSEF via 
eSSC at POSTECH.
\end{acknowledgments}

\end{document}